\documentclass{aa}  

\usepackage{graphicx}
\usepackage{txfonts}
\usepackage{color}

\def\arcsec{$^{\prime\prime}$}
\begin{document}

   \title{The active region source of a type III radio storm observed by Parker Solar Probe during Encounter 2}

\author{L. Harra
          \inst{1,2}
          \and D. H. Brooks\inst{3}
          \and S. D. Bale \inst{4}
           \and C. H.  Mandrini \inst{5,6}
          \and K. Barczynski \inst{1,2}
          \and R. Sharma \inst{7}
          \and S. T. Badman \inst{4}
          \and S. Vargas Dom\'inguez \inst{8}
          \and M. Pulupa \inst{4}
    }      

   \institute{PMOD/WRC, Dorfstrasse 33
CH-7260 Davos Dorf, Switzerland\\
              \email{louise.harra@pmodwrc.ch}\\
              \email{krzysztof.barczynski@pmodwrc.ch}
         \and
         ETH-Zurich, H\"onggerberg campus, HIT building, Z\"urich, Switzerland
        \and 
             College of Science, George Mason University, 4400 University Drive, Fairfax, VA 22030 USA\\
            \email{dhbrooks.work@gmail.com}
             \and 
             Physics Department and Space Sciences Laboratory, University of California, Berkeley, USA. 94720-7450\\
           \email{bale@berkeley.edu}
           \and
           Instituto de Astronom\'ia y F\'isica del Espacio (IAFE), CONICET-UBA, Buenos Aires, Argentina 
           \and 
           Facultad de Ciencias Exactas y Naturales (FCEN), UBA, Buenos Aires, Argentina\\
           \email{mandrini@iafe.uba.ar}
           \and Fachhochschule Nordwestschweiz (FHNW), Bahnhofstrasse 6, 5210 Windisch, Switzerland \\
           \email{rohit.sharma@fhnw.ch}
           \and
           Universidad Nacional de Colombia, Observatorio Astronómico Nacional, Bogot\'a, Colombia \\
           \email{svargasd@unal.edu.co}
          } 
         %    \thanks{}

   \date{Received September 2020 }

% \abstract{}{}{}{}{} 
% 5 {} token are mandatory
 
  \abstract
   {To investigate the source of a type III radio burst storm during encounter 2 of NASA's Parker Solar Probe (PSP) mission. }
   {It was observed that in encounter 2 of NASA's Parker Solar Probe mission there was a large amount of radio activity, and in particular a noise storm of frequent, small type III bursts from 31st March to 6th April 2019. 
   %%were frequent and small type III bursts over a number of days. 
   Our aim is to investigate 
   %%determine 
   the source of these small and frequent bursts.}
   {In order to do this, we analysed data from the Hinode EUV Imaging Spectrometer (EIS), PSP FIELDS, and the Solar Dynamics Observatory (SDO) Atmospheric Imaging Assembly (AIA). We studied the behaviour of active region 12737, whose emergence and evolution coincides with the timing of the radio noise storm
   %%which is the only viable source,
   and determined the possible origins of the electron beams within the active region.
   %%where the SEPs originate. 
   To do this, we probe the dynamics, Doppler velocity, non-thermal velocity, FIP bias, densities, and carry out magnetic modelling. }
   {We demonstrate that although the active region on the disk produces no significant flares, its evolution %and other smaller-scale dynamics 
   indicates it is a source of the electron beams causing the radio storm.
   %SEPs.
   They most likely originate from the area at the edge of the active region that shows strong blue-shifted plasma. We demonstrate that as the active region grows and expands, the area of the blue-shifted region at the edge increases, which is also consistent with the increasing area where large-scale or expanding magnetic field lines from our modelling are anchored. This expansion is most significant between 1 and 4 April 2019, coinciding with the onset of the type III storm and the decrease of the individual burst's peak frequency, indicating the height at which the peak radiation is emitted increases as the active region evolves.}
   % when we also see an order of magnitude increase in the strength of the type III bursts, and an 
 %  This also occurs at the same time as an increase in the first ionization potential (FIP) bias. This increase is consistent with an increase in Solar Energetic Particles \citep{Reames2018}}.
   
   %In addition, we find that the blue-shifted region shows the most persistent variability, consistent with the persistence and minute timescale repetition rate of the type III storm at this time. }
   %The type III bursts are always seen in PSP during this time period, and hence a physical source that produces persistent variability is required.   }


   \maketitle
%
%-------------------------------------------------------------------

\section{Introduction}

  Type III radio bursts are observed regularly on the Sun, and their sources are beams of energetic electrons streaming outwards along open magnetic field lines. The largest Solar Energetic Particle (SEP) events are due to shocks associated with coronal mass ejections (CME) and solar flares e.g. \cite{reames}, though others can be associated with jets.  For example, \cite{krucker} made a study of jets associated with flares as sources of supra-thermal electron beams. The flares in this paper all had energies higher than C2 GOES classification.  However, electrons can be accelerated in smaller energy release events in active regions,  when magnetic field in a coronal hole interacts with nearby closed magnetic field, and even in bright points that produce jets \citep[see the review by][]{hamish}. 
  
NASA's  Parker Solar Probe (PSP) mission has opened up a new way of probing type III bursts due to the close vicinity to the Sun. \cite{pulupa} have explored the statistics of type III bursts during the first two encounters of PSP. They found that only a few bursts occurred during encounter 1 (E01, October-November 2018, perihelion November 6th 2018), during which there was minimal solar activity. In encounter 2 (E02, March-April 2019, perihelion April 4th 2019), however, there were a number of dynamic active regions, and a large number of type III radio bursts observed by PSP. In particular, two active regions (AR 12737 and AR 12738) were prominent during this time interval and at solar locations such that radio waves would be likely to reach PSP. AR 12738 was the larger active region, receiving its NOAA designation on April 6 as it rotated onto the solar disk as viewed from Earth, but had existed at least 2 weeks prior and was visible in STEREO A/SECCHI data. \citet{Krupar2020} used radio triangulation of a single strong radio burst on April 3 and showed its location to be consistent with a Parker spiral emerging from AR 12738, indicating it was responsible for at least some of the radio activity observed at this time. In addition, Cattell et al. (this issue), find some evidence of correlating periodicities in AR 12738 and in radio at PSP, although the interval they investigated is well after the time period studied in this paper, once AR 12738 had rotated on disk and AR 12737 had rotated off disk.

AR 12737, meanwhile, was observed to develop out of a coronal bright point near the East limb on March 31 and evolve and grow before reaching a more steady state on approximately April 6th. At this time, in addition to strong impulsive bursts which PSP measured throughout this encounter and which \citet{Krupar2020} were able to associate with the larger active region, a significant type III radio noise storm occurred consisting of a huge number of smaller and more quickly repeating radio bursts. In addition, this noise storm showed the interesting feature of a systematic decrease in peak frequency with time at the exact time that AR 12737 was emerging and developing. This will be discussed further in relation to Figure \ref{type3_all}, but here we simply state that for the above reasons, it is well motivated to study this active region in relation to the noise storm as this time. In addition, in contrast to the larger active region, AR 12737 is visible on disk during PSP's closest approach to the Sun (April 4th), and thus more likely to be measuring weaker radio events which are harder to study at 1AU. Hence AR 12737 is observed at this time with all the available Earth-based instrumentation including imaging spectroscopy, radio interferometry, magnetographs and more. Since the active region is not flare active during our period of interest, we consider other possible sources of the type III bursts. Other possibilities are jets, micro-flares and active region outflows.
%They found that only a few bursts occurred during encounter 1 (E01). At this time, PSP was connected to a corona hole. In encounter 2 (E02), there is an active region (12737) on the disk, that is not flare active, with a larger region (12738) appearing around the west limb on 6 April 2019. Based on modelling of the connectivity to the Sun \citep[see ][for a description of the methodology for E01]{Badman2020}, PSP was not connected to either active region. The co-rotation loop for PSP at perihelion (April 4 -- at a distance of $\approx$ 35.7 $R_{\odot}$) suggests a connection $\sim$30 degrees to the solar East side of AR 12737 between 1 and 4 April. Of course, a direct magnetic connection is not necessary for PSP to detect type III bursts. {\bf Stuart - add something in about the spread of SEPs}

Jets are an important aspect of electron acceleration, and indeed have been put forward as a means to explain the magnetic switchbacks that have been seen so clearly by PSP \citep{Bale2019,2019Natur.576..228K,2020ApJS..246...45H}. \cite{alphonse} have proposed that reconnected minifilament eruptions can manifest as outward propagating Alfv\'enic fluctuations that steepen into a disturbance as they move through the solar wind.
\cite{Mulay_Sharma_2019} showed a spatial association of active region jet and the interplanetary type-III burst source co-spatial with extrapolated open magnetic field lines. The bunches of type-III bursts occurred before and during jet eruption suggesting particle acceleration begins before the EUV jet eruption. 

Another potential source of type III bursts are the blue-shifted regions found at the edges of active regions \citep[e.g.,][]{flows}. The blue-shifted regions have been put forward as one of the potential sources of the slow solar wind, and they have been found to often have elemental abundances consistent with the slow solar wind \citep[e.g.,][]{Brooks2011}. \cite{DelZanna2011} have also studied these outflows and suggest that the continuous growth of active regions maintains a steady reconnection across the separatrices at the null point in the corona. The acceleration of electrons in the interchange reconnection region between the closed loops in the AR core and the outflow on open field lines at the boundary, produces a radio noise storm in the closed loop areas, as well as weak type III emission along the open field lines. 

In this paper, we explore the data from the second solar encounter of PSP (E02) around perihelion on 4 April 2019, and investigate the possible source(s) associated with AR 12737 of the type III radio storm that occurs in the lead up to this time.
%bursts from AR 12737. 
% \cite{pulupa} suggest that the most likely origin of the type III bursts in E02 is active region NOAA 12738, but this refers to the later time period when the AR appeared on disk. The modelled footpoints of PSP do not track nearer to AR 12738 until after April 14. 

%--------------------------------------------------------------------
\section{Instrumentation and data analysis}

The Hinode EUV Imaging Spectrometer (EIS) instrument \citep{eis} is an imaging spectrometer that has two narrow slits that raster to build up images, and two slots. In this work we use studies with the 2\arcsec slit, where spectral images have been built up that cover the whole active region starting from 1 April less than a day after the region first emerged. In addition, there is a fast 3-step raster with a cadence of 42 seconds on 1 April starting at 17:00 UT lasting an hour. The field of view in this case covers a strip towards the eastern edge of the active region. We processed the EIS data using the standard calibration procedure eis$\_$prep. The emission lines were fitted with single or multiple Gaussian profiles (depending on the presence of known blends) in order to extract plasma parameters such as Doppler velocity, non-thermal velocity, and FIP (first ionization potential) bias. 

The spectroscopic data were combined with Atmospheric Imaging Assembly (AIA) data from the \emph{Solar Dynamics Observatory} \citep[SDO;][]{Pesnell2012}  to provide context imaging and an understanding of the dynamical behaviour of the action region as it emerged on 31st March and developed through to 7th April 2019. 

Measurements of interplanetary type III radio bursts were made by the Radio Frequency Spectrometer (RFS) subsystem \citep{2017JGRA..122.2836P} of the FIELDS instrument suite \citep{2016SSRv..204...49B} on the NASA Parker Solar Probe (PSP) mission \citep{2016SSRv..204....7F}.  The FIELDS/RFS system is a base-band radio receiver that produces full Stokes parameters in the range 10.5 kHz - 19.17 MHz, corresponding to plasma frequencies at radial distances of $\sim$ 1.6 $R_S$ to 1 AU \citep{1998SoPh..183..165L}.  Voltage measurements are made on two $\sim$ 7m crossed dipoles and digitized into 2 virtual receivers (the High Frequency Receiver - HFR and the Low Frequency Receiver - LFR) in 128 pseudo-log spaced ($\Delta f/f \approx 4\%$) frequency bins with one spectrum each 7 seconds during normal solar encounter operations.  Auto- and cross-spectra are used to produce Stokes parameters.  The RFS system has good sensitivity down to the level of the galactic synchrotron spectrum \citep{pulupa} and so is capable of measuring very weak solar radio bursts.

Solar observations using the Murchison Widefield Array (MWA) were made during the period of Parker Solar Probe (PSP) perihelion. However, the observations were corrupted by incorrect pointing, except on 5th April. The Murchison Widefield Array (MWA) located in Western Australia is a new generation low-frequency radio interferometer based on large-'N' design \citep{Lonsdale2009,Tingay2013}. It has 128 elements and operates between the 80 MHz to 300 MHz frequency range. Here, we use MWA solar data from the phase-II configuration, with baselines extending up to 5 km \citep{MWA_Phase2}. Long baselines, accompanied by a large number of elements, make MWA suitable for studying small but extended features like active regions. 
 We analyse 5 minutes of MWA solar data from 5$^{th}$ April 2019 04:00:16 to 04:05:16 UT taken during the G0002 solar observation program. The solar observations were made in `picket-fence'  mode, i.e. the total available bandwidth of 30.72 MHz can be distributed within 80-300 MHz in 12  roughly log-spaced chunks,  each of 2.56 MHz wide. Here, we analyse 4 frequency bands at 107.5, 163.8, 192.0, 240.6 MHz. 

Since our aim is to find the origin of the type III noise storm measured at PSP from 1 to 6  April 2019, we have modelled the coronal magnetic field of AR 12737 during that period in search of large-scale or expanding field lines (i.e. 'open' within the limitations of a local magnetic field model approach) that could be associated to the areas where EIS observes blue-shifted regions. Our model is carried out by taking as boundary condition the vertical component of the photospheric magnetic field derived from SDO/HMI observations. The magnetic field vertical component is computed from the line-of-sight magnetograms downloaded from the Joint Science Operation Center (JSOC, http://jsoc.stanford.edu/HMI/Magnetograms.html). We selected data with a 720 ms resolution from that database.

\section{A type III storm and the active region}
\label{section3}
During PSP encounter 2, frequent type III radio bursts were observed as described by \cite{pulupa}. We concentrate on the period between 31 March 2019 - 6 April 2019. 

\begin{figure*}
   \centering
   \includegraphics[width=0.9\textwidth]{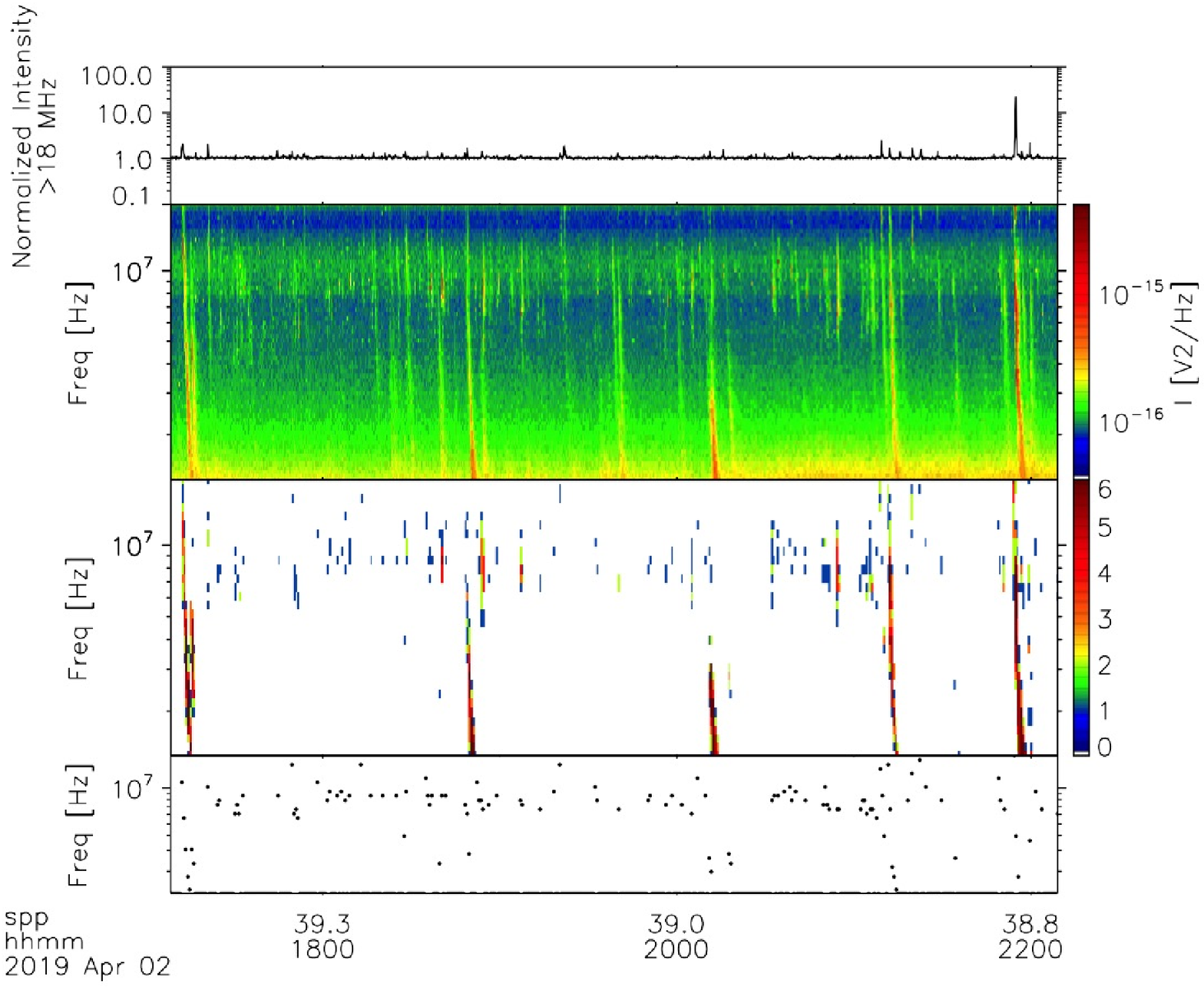}
   \caption{PSP/RFS radio frequency data showing the type III bursts and storm on April 2, 2019.  Top panel is the normalized Stokes intensity above 18 MHz, 2nd panel is the full spectrogram, 3rd panel shows significant burst above background, and the bottom panel is the frequency of maximum normalized intensity.  This interval shows both more classical type III bursts, and the weaker more localized features associated with type III radio storms.  These data are described further in the text.}
    \label{type3_zoom}%
\end{figure*}

Figure~\ref{type3_zoom} shows PSP/RFS data during a $\sim$5 hour time period on April 2, 2019.  The top panel is the radio Stokes intensity $I$ in a 40 second window normalized to the mode (most probable) value in that window, at frequencies 18.28-19.17 MHz (the highest two frequency bins).  The 2nd panel is the full RFS/HFR Stokes $I$ spectogram, at full spectral and temporal resolution.  In these panels, one can see both larger intensity type III bursts, extending over the full frequency band and exhibiting the classical frequency drift, {\em and} the more impulsive, frequency-localized features associated with type III radio storms.  The third panel shows, as intensity, the number of 7-second intensity measurements that exceed 2$\times$ the mode value (as per the top panel) in a 40 second window (hence the maximum value is 5), as a function of measurement frequency.  This shows the location in time-frequency of `bursts' rather than background (galactic) noise.  The bottom panel is the frequency of the maximum normalized intensity (in the panel above).  While not shown in Figure~\ref{type3_zoom}, Stokes linear polarization U/I and Q/I are measured for this type III storm to be less than \%2.  While strong linear, or circular, polarization can be an indication of mode conversion processes \citep[e.g.][]{1981A&A...101..250Z} and, therefore, a clue to emission mode (fundamental or harmonic), the relatively weak polarization signature observed here probably indicates strong scattering in density fluctuations.  We take it as an indication of fundamental emission.

\begin{figure*}
   \centering
   \includegraphics[width=0.95\textwidth]{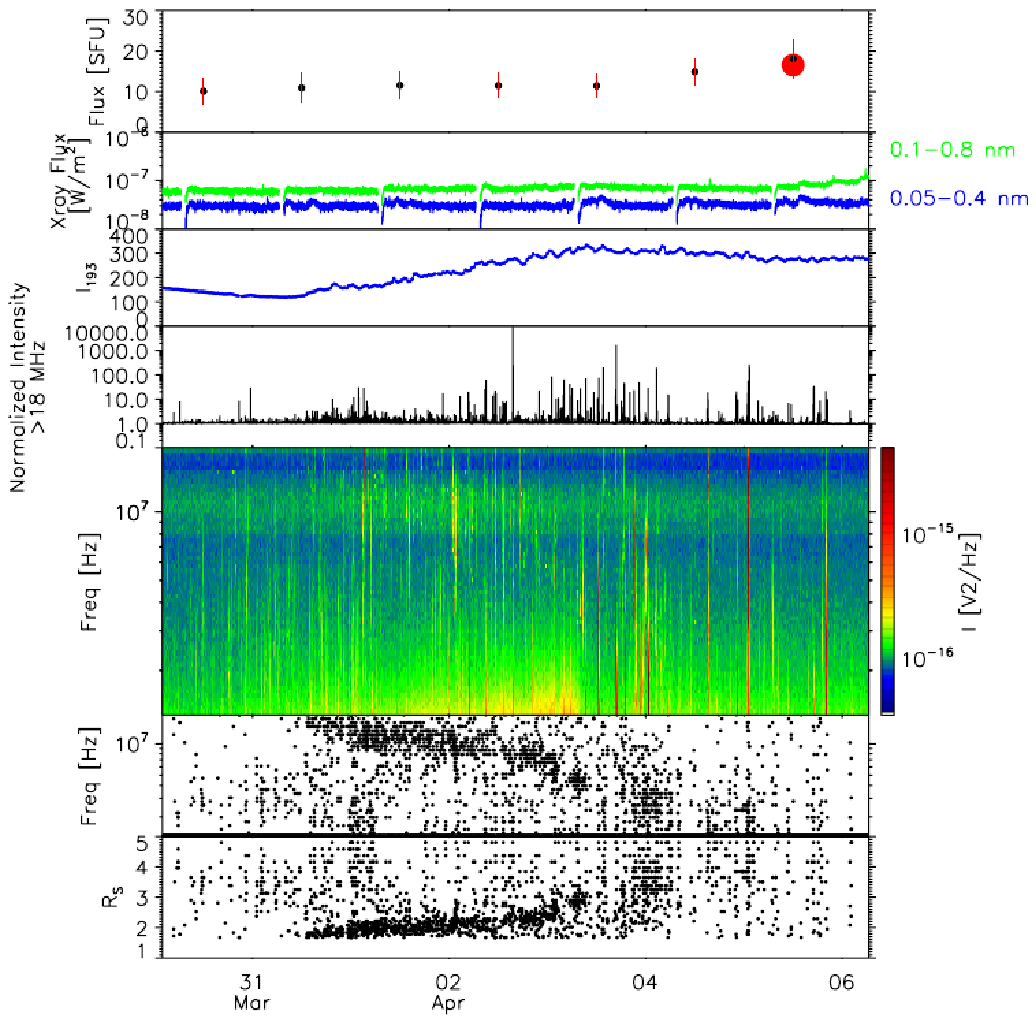}
   \caption{PSP/RFS, RSTN, and MWA radio frequency measurements, AIA 193 \AA ~intensity, and GOES X-ray data over the full interval March 30, 2019 - April 6, 2019.  The top panel shows daily RSTN flux from 30th March to 6th April 2019 at 245 MHz. The red circle marks the flux at 240 MHz obtained from MWA flux calibration, with error bars from the temporal RMS each day.  The second panel is GOES soft X-ray flux data which shows no significant flares.  The 3rd panel shows the AIA 193 \AA ~intensity integrated over the FOV shown in Figure~\ref{ar_evolution}, showing the growth and flattening off of the development of the AR from 31 March through 4th April.  The 4th panel 
   is again the normalized Stokes intensity above 18 MHz and 5th panel is again the full spectrogram.  The 
   6th panel is the frequency of peak normalized signal (as per Figure~\ref{type3_zoom}).  The bottom panel 
   is the inferred source height of the radio emission, inverting the peak frequency from a solar wind density profile and fundamental emission, as described in the text.  The noise storm commences on March 31 and a clear trend between April 1 to April 4 shows the peak type III storm emission frequency decreasing, corresponding to a higher source altitude or a more rarefied source region.}
    \label{type3_all}%
\end{figure*}

Furthermore, as seen in Figure~\ref{type3_all}, Panel 6, the frequency of maximum radio intensity is decreasing with time over this interval.  As shown in Panel 7, this implies a source moving to higher altitudes (and hence lower heliospheric plasma density) or an expanding source region whose density decreases with lateral expansion.  The lower panel in Figure~\ref{type3_all} shows the inferred height from a heliospheric density profile $n_e(r)$ \citep{1998SoPh..183..165L}, assuming radio emission at the fundamental $f_{pe}$ as is commonly assumed for type III radio storms \citep{morioka2015}.

\begin{figure*}
   \centering
   \includegraphics{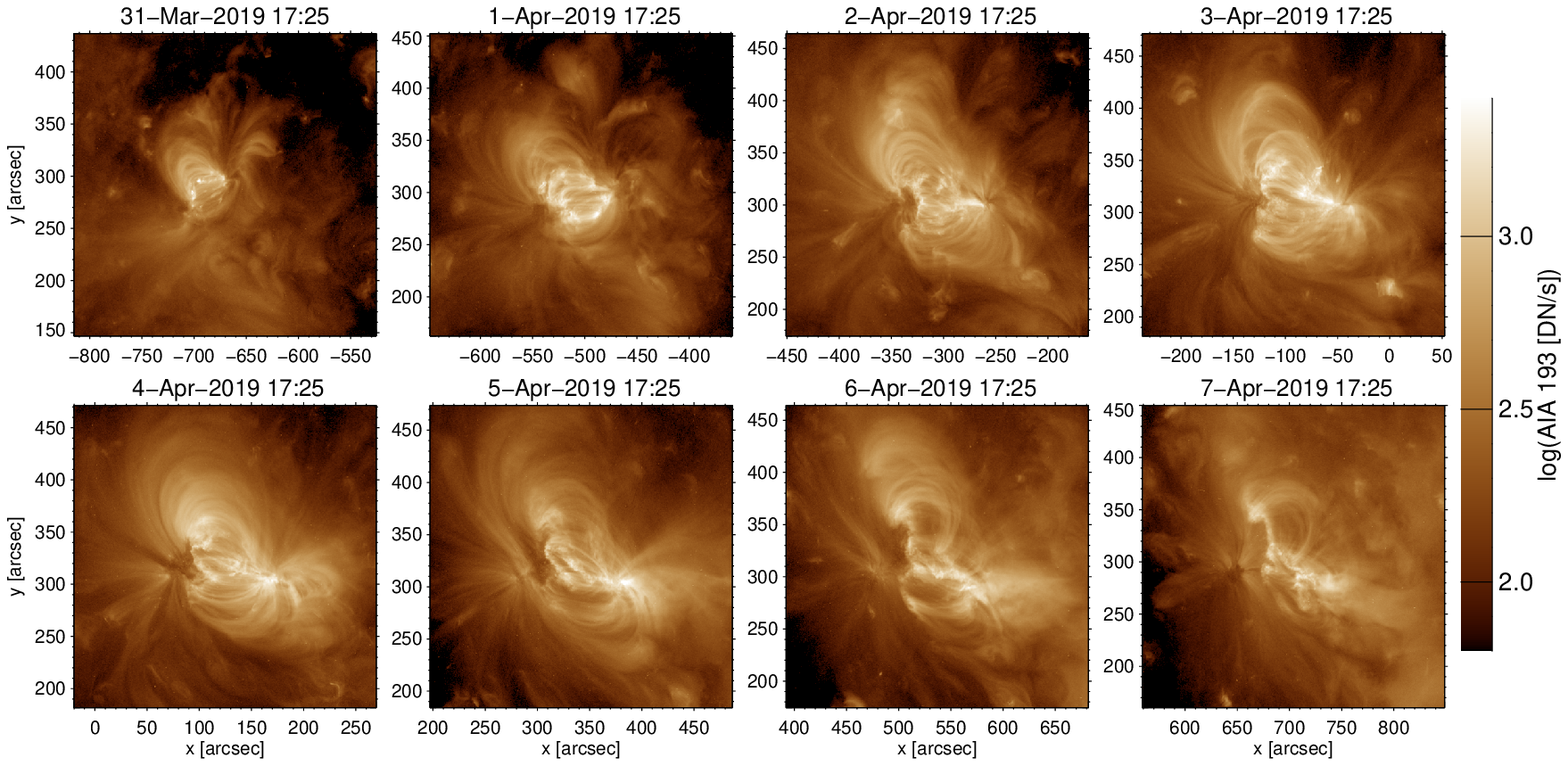}
   \caption{The evolution of the active region from its emergence on the 31st March to 7 April 2019. The active region has a simple structure, with expansion clearly seen, especially between the 31 March and 3 April.  }
              \label{ar_evolution}%
    \end{figure*}

During this time period, $AR 12737$ can be seen to emerge in the corona near the east limb (Earth-referenced). As the region develops (see Figure \ref{ar_evolution}), it's overall extreme ultraviolet (EUV) emission increases (panel 3 in Figure \ref{type3_all}). This onset and following transient change coincides with the onset of the noise storm picked out by the peak frequencies in panel 6 of Figure~\ref{type3_all}. The frequency (or corresponding height of emission) change monotonically at the same time as the EUV emission increases monotonically. Once the EUV emission levels out the active region becomes more stable, the type III storm is observed to level out and gradually dissipate.

% From 31 March to the 4th April there is only one active region on the solar disk. The progression of the active region increases the mean emission levels at radio wavelengths. 
In addition, RSTN\footnote{\texttt{https://www.sws.bom.gov.au/World\_Data\_Centre}} monitored the solar radio flux and recorded an increase towards the end of the week. The top panel of Figure \ref{type3_all} plots the daily-averaged solar radio flux for 245 MHz. A distinct increase can be seen from 4$^{th}$ April, which is due dominantly to  AR 12737 as it is the only region on the disk at that time.  We note that AR 12738 is also visible from April 6, and may contribute to the increasing RSTN flux after that date \citep{Krupar2020}.
 %The RSTN flux is sky-integrated. However, the change from 3rd April to 5th April is apparent in radio flux. During these three dates the active region from eastern limb is not visible. In addition, on 5th April if that active region was bright, then we would have seen in the MWA radio maps. During these days, the rising active region is not brighter than on-disk active region.

%

%\begin{figure*}
%    \centering
%    \includegraphics[width=0.7\textwidth]{RSTN_flux1.png}
%    \caption{Daily RSTN flux from 30th March to 7th April 2019 at 245 MHz. The red square marks the flux at 240 MHz obtained from MWA flux %calibration. The error bars come from the temporal RMS from each day.}
%    \label{fig:rstn}
%\end{figure*}
% As discussed in the Introduction, the AR is not magnetically connected to PSP but SEPs can spread in longitude and be detected by unconnected spacecraft.  
During E02, PSP does not observe the type III electron beams \textit{in situ} causing the radio emission. This is not surprising since simple ballistic modeling and a PFSS model \citep[see][ for details on method]{Badman2020} shows that PSP did not connect magnetically to either active region during this time. Such a connectivity would make the origin of the type III bursts less ambiguous but unfortunately is not possible in this case. Type III radio emission is widely beamed, and bursts can often be seen by widely separated spacecraft, and thus proximity to one active region or another is not a strong constraint on the source location. As such, it is plausible for PSP to be receiving radio emission from electron beams injected by AR 12737 and so we proceed with a full analysis of its dynamics and properties to determine its nature as a radio source.

Our goal is to understand where the source of the SEPs could be within this active region. The first check is to see if there are any flares, but the active region produces no flares greater than GOES `A' level (see second panel, Figure \ref{type3_all}). There are no significant energy release events in this time period. 

%-------------------------------------- Two column figure (place early!)

   The active region emerges as a simple bipole on 31 March 2019. Figure ~\ref{ar_evolution} shows AIA images from its emergence until 7 April 2019. In the first two time-frames the active region is compact. By 3 April, there are many more bright loops, covering a larger area, and the edges of the active region are showing extended structures. In the same time period,
   % the amplitudes of the types III bursts have increased by at least an order of magnitude. 
   as shown in Figure \ref{type3_all}, the type III storm of interest commences and shows a transient and monotonic change in its peak frequency (implying source evolution), at the same time as the active region evolves.
    
\begin{figure*}
   \centering
   \includegraphics[scale=0.4]{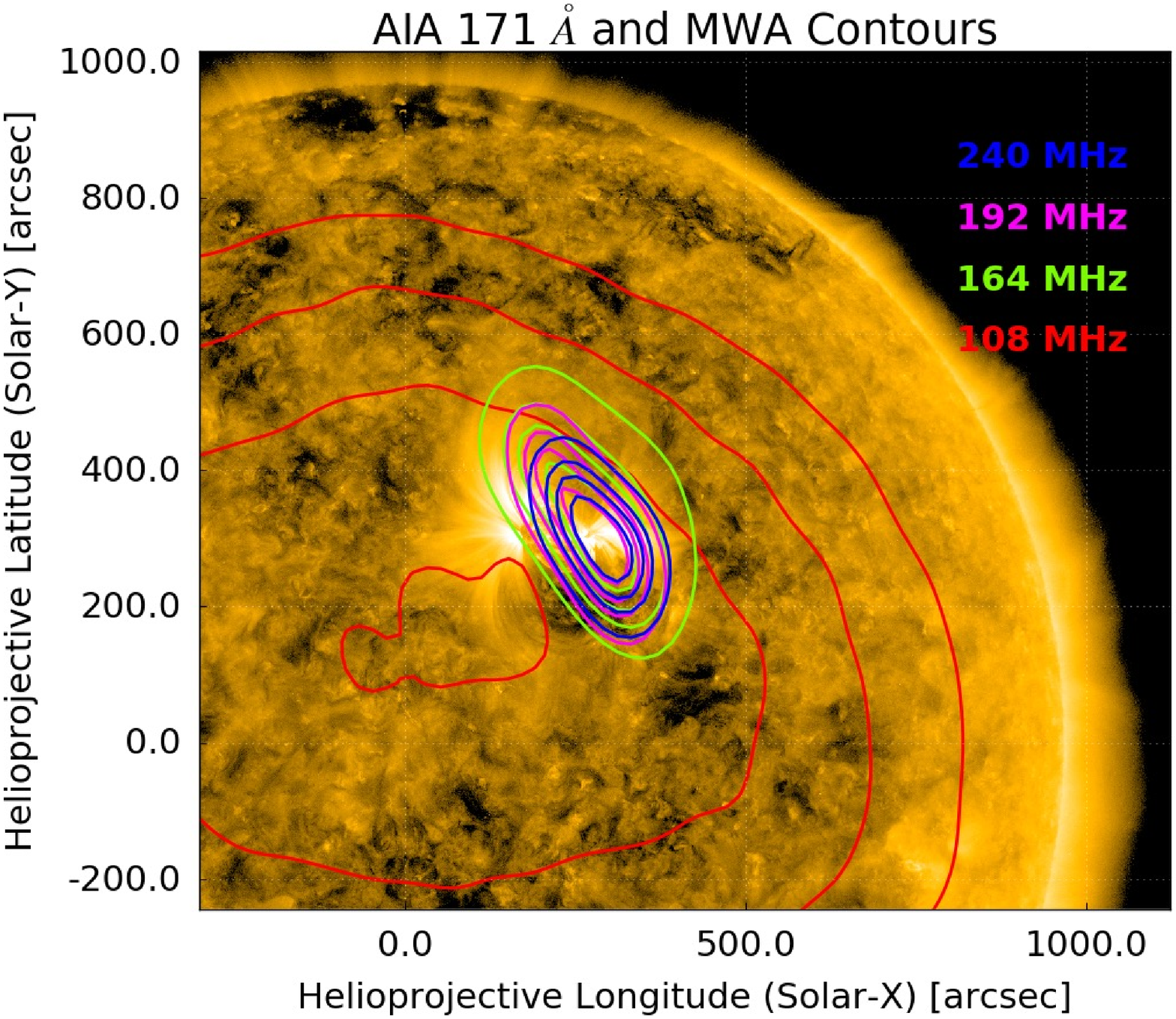}
   \caption{AIA's 171 \AA \ solar image overlaid with the MWA radio contours at 4 frequency bands. The time corresponding to the AIA image is 04:01:45 UT. The contour levels corresponding to 164, 192 and 240 MHz are 50\%, 70\%, 80\% and 90\% w.r.t map's maximum T$_{B}$ respectively. The contour levels corresponding to the 108 MHz radio source are 70\%, 80\%, 90\%, 99\%, w.r.t map's maximum T$_{B}$.  }
              \label{radio}%
    \end{figure*}

 To determine if the AR is a radio source, we produced 2.56 MHz bandwidth and 10s time-averaged MWA solar maps. The solar emission in the image was defined by choosing a 5-$\sigma$ threshold, where $\sigma$ is the RMS noise computed over the region far from the Sun in the image. The resultant MWA radio images were converted into brightness temperature (T$_{B}$) maps following the procedure described in \cite{Mohan2017}. The contour plots of the T$_B$ maps are shown in Figure \ref{radio} for the 4 frequency bands. The location of the contours suggests radio sources at 162 MHz, 192 MHz and 240 MHz are associated with the active region. No other radio source was observed in the MWA solar maps.
 %The location of the contours for 108 MHz is shifted eastwards from the active region. The radio source at 162 MHz and 192 MHz also shows a westward extension at low-contour levels, i.e. at 10\% and 20\% level w.r.t the peak.
  
\begin{figure*}
   \centering
   \includegraphics[scale=0.6]{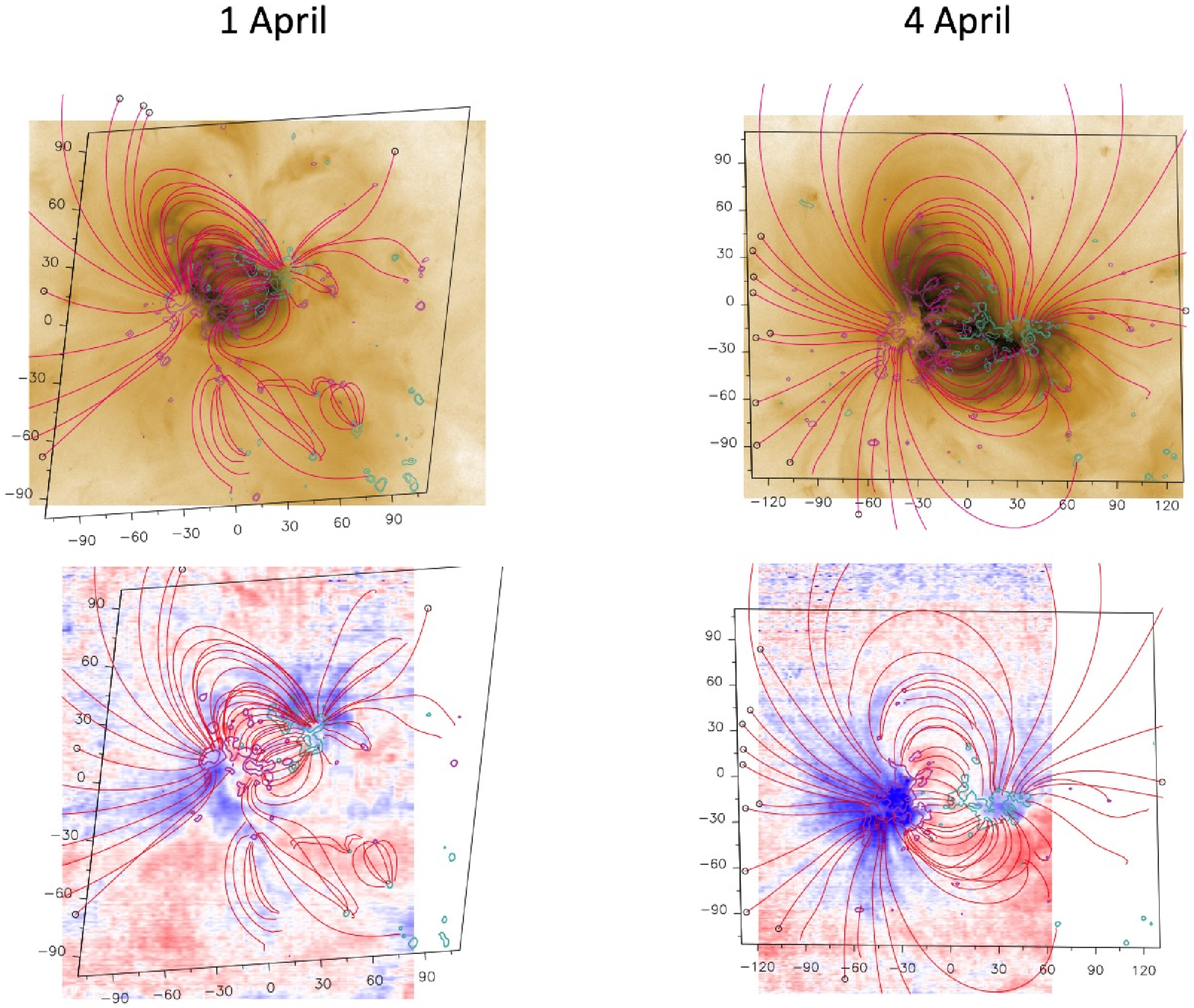}
   \caption{ The top two panels show the magnetic field model of the active region on 1 April and 4 April overlaid on AIA  193~\AA~images with the intensity reversed. The bottom two panels show the same model but overlaid on the Doppler velocity maps with blue showing blue shifts and red showing red shifts (colour range is +/-20km/s).  The AR expands as it evolves. At its edges on 1 April, we observe structures in AIA data that are closed within the AR. On 4 April the magnetic field lines derived from our model look more expanded at the edges of the AR, and larger scale magnetic 
   field lines are relevant in correspondence with the blue-shifted region. The axes in all panels are labelled in Mm, with the origin set in the AR centre (located at N12E31 on 1 April and at N13W05 on 4 April). The iso-contours of the line-of-sight field correspond to $\pm$ 50, $\pm$ 100 and $\pm$ 500 G in continuous magenta (blue) style for the positive (negative) values. Sets of computed field lines matching the global shape of the observed coronal loops in the AIA 193~\AA~images have been added in continuous line and red colour in the two top panels and repeated in the two bottom panels.}
              \label{magnetic_model}%x
    \end{figure*}

     Figure~\ref{magnetic_model} shows the global coronal structure derived from our modeling for 1 and 4 April for which we have used the closest in time magnetograms to each EIS observation. The figure is constructed superposing field lines computed using different linear-force-free field (LFFF) 
models, i.e. $\vec{\nabla} \times \vec{B}= \alpha \vec{B}$, where $\vec{B}$ is the magnetic field and $\alpha$ is a constant \citep[see][and references therein for a description of the model and its limitations]{Mandrini15}. Because we are interested in the large scale coronal configuration and the potential presence of `open' field lines, we selected a region four times larger than that covered by AR 12737, and centered on it, for each model. Each of these magnetic maps was embedded within a region twice as large padded with a null vertical field component to, on one hand, decrease the modification of the magnetic field values since the method to model the coronal field imposes flux balance on the full photospheric boundary (i.e. flux unbalance is uniformly spread on a larger area and the removed uniform field is weaker as the area is larger), and, on the other, to decrease aliasing effects resulting from the periodic boundary conditions used on the lateral boundaries of the  coronal volume. Doing so, we are able to discriminate field lines that connect to the surrounding quiet-Sun regions from those that are potentially 'open' magnetic field lines as they leave the extrapolation box. In particular, field lines ending in an open circle in Figure~\ref{magnetic_model} are those that leave the  computational box shown in each panel. In all our models the height above the photospheric boundary was 400 Mm.

The free parameter of each of our LFFF models, $\alpha$, has been selected to better match the shape of the observed loops in AIA 193~\AA~(see Figure~\ref{ar_evolution}). To do this comparison, the model is first transformed from the local frame to the observed frame as discussed in \citet{Mandrini15} \citep[see the transformation equations in the Appendix of][]{Demoulin97}. This allows a direct comparison of our computed coronal field configuration to AIA EUV images obtained at almost the same time and shown as background in Figure~\ref{magnetic_model}. Furthermore, in order to determine the best matching $\alpha$ values we have followed the procedure discussed by \citet{Green02}. AR 12737 is a mainly bipolar AR that emerges close to the eastern solar limb around 31 March; it expands and decays as it evolves on the solar disc, until it disappears on the western limb on 7 April. During this period only a few minor flux emergence episodes are seen. The two bottom panels in Figure~\ref{magnetic_model} illustrate very well that the blue-shifted region is mostly consistent with large-scale or expanding magnetic field lines whereas the red-shifted region is mostly consistent with closed loops.

%In general its coronal structure is indicative of a negative magnetic helicity sign as would correspond to a northern AR according to the hemispheric AR rule \citep{Pevtsov03}. Consequently, the values of $\alpha$ found for the 1 and 4 April models are all negative. To match the best possible the observed AIA loops on 1 and 4 April, we used two different values of $\alpha$; $-1.1 \times 10^{-2}$ Mm$^{-1}$ and $-9.4 \times 10^{-3}$ Mm$^{-1}$ for 1 April, while for 4 April the values were $-9.4 \times 10^{-3}$ Mm$^{-1}$ and $-3.1 \times 10^{-3}$ Mm$^{-1}$. In both cases, the most negative values correspond to the close field lines in the northern part of the AR, while the less negative ones to the southern part and the largest scale loops.        

\begin{figure*}
   \centering
   \includegraphics[scale=0.4]{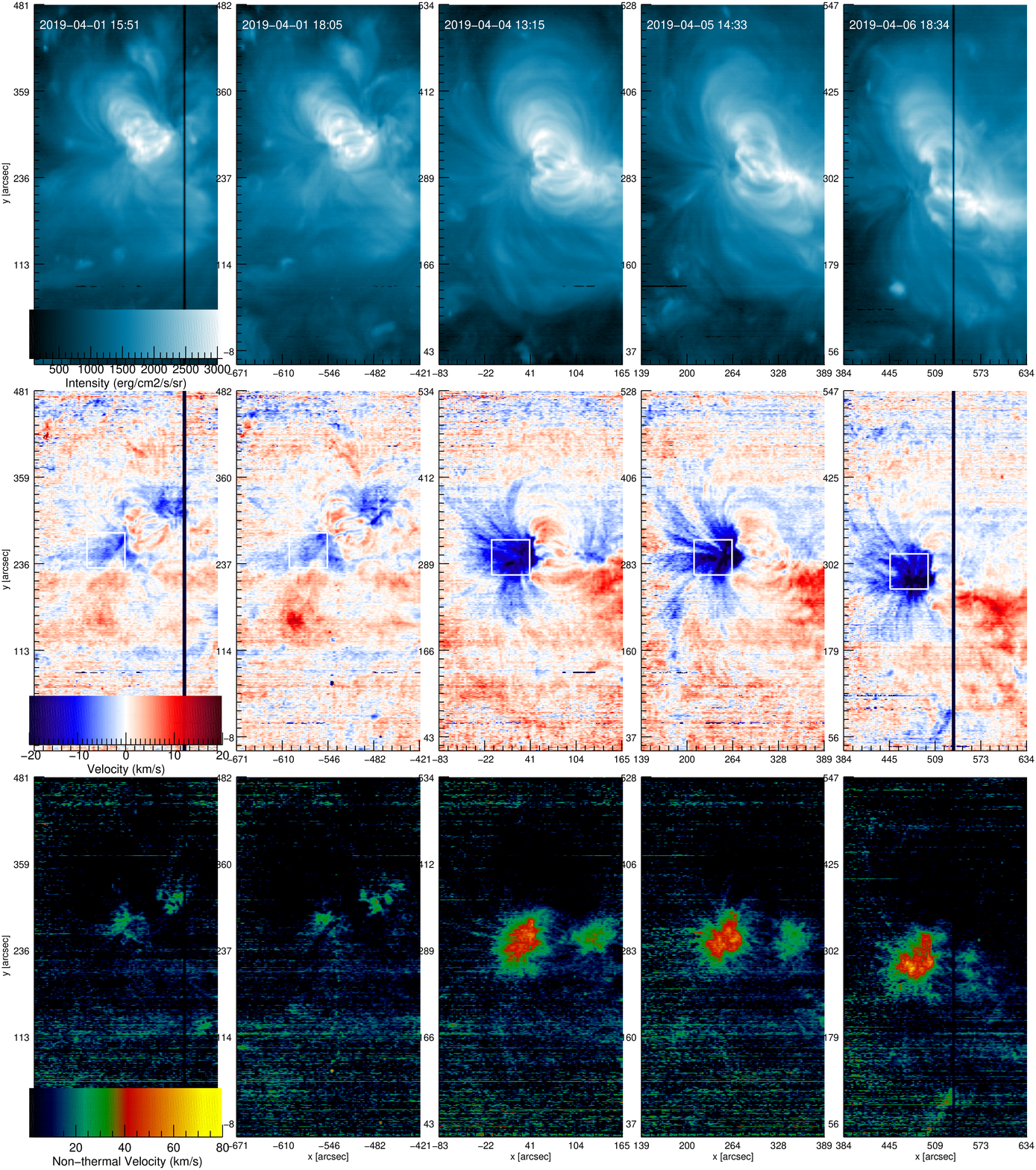}         
   \caption{Top panel: EIS Fe {\sc xii} 195.119\,\AA\, intensity maps of the active region as it crossed the disk from 1 -- 6 April. Centre panel: Doppler 
velocity maps derived from the same line. The colour bar shows the velocity range. The white boxes indicate the outflow regions chosen for measuring the FIP bias. 
Bottom panel: Non-thermal velocity maps also from Fe {\sc xii} 195.119\,\AA. }
              \label{FIP_boxes}%
    \end{figure*}

    Figure~\ref{FIP_boxes} shows EIS Fe {\sc xii} 195.119\,\AA\, intensity images, and Doppler and non-thermal velocity maps from 1 -- 6 April. The Doppler velocity has a strong radial component whereas the non-thermal velocity does not show this \cite{mariska}. This makes the non-thermal velocity a location-independent measure.
The standard EIS data processing uses an artificial neural network model to determine the orbital drift of the spectrum on the CCD due to thermal
instrumental effects \citep{kamio2010}. This is an important step in producing accurate velocities since EIS does not have an absolute wavelength calibration.
The maps in Figure~\ref{FIP_boxes} therefore show Doppler velocities relative to a chosen reference wavelength. In this case we used the mean centroid in the
top part of the raster FOV. The neural network model also utilizes instrument housekeeping information from early in the EIS mission, so the correction is less
applicable to more recent data and often leaves a residual orbital variation across the raster. We therefore corrected the velocities in a final step by 
modeling this residual effect \citep[see Appendix of ][]{Brooks2020}. For the non-thermal velocity calculation we followed the procedure outlined by
\citep{Brooks2016}. 

\begin{figure*}
   \centering
   \includegraphics[scale=0.6]{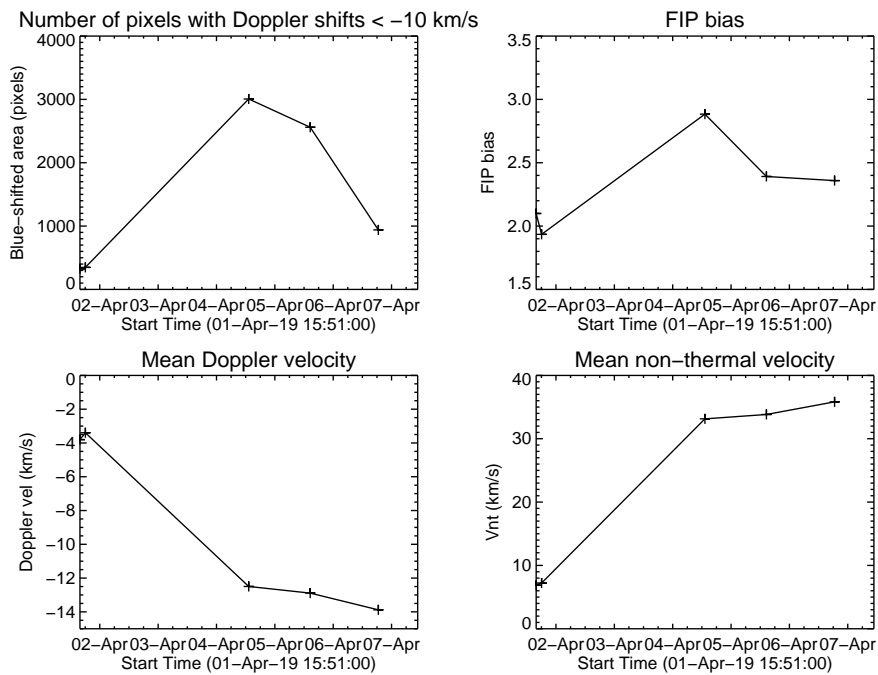}
   \caption{Top left panel: number of pixels of the blue-shifted region as the active region crossed the disk from 1 -- 6 April. The area of the blue-shifted region increases as the AR evolves. Top right panel: FIP bias measurements in the outflow area (white boxes in Figure \ref{FIP_boxes}). The
FIP bias appears to increase on April 4th when the blue-shifted area also reaches a maximum. Bottom left panel: Variation of the mean of Doppler velocity in the same region. Bottom right panel: Variation of mean of non-thermal velocity in the same region. Both the Doppler velocity and non-thermal velocity show a significant increase in magnitude in the first few days of the active region's formation. }
              \label{FIP}%
    \end{figure*}

The AR grows and expands during this period. Although there are no rasters between 1 April and 4 April, we can clearly see that between these two dates the area of the blue-shifted region to the eastern side of the active region is larger. The non-thermal velocities in the same region also expand in area and increase in magnitude.  This is consistent with the images in Figure~\ref{ar_evolution} which show clear expansion of the active region. The blue-shifted region also coincides with large-scale or expanding magnetic field lines from our modelling (Figure~\ref{magnetic_model}), so plasma can find a way to escape into the solar wind. We examined the area of the blue-shifted region more quantitatively in Figure~\ref{FIP} (top left panel). It shows the number of pixels in the blue-shifted region. This area of blue-shifted plasma increases as the AR evolves. 

We also measured the FIP bias in the outflow region (highlighted by white boxes) as it crossed the disk (Figure~\ref{FIP}; top right panel). 
The FIP bias is enhanced by about a factor of 2 above photospheric abundances throughout the period of the observations, but appears to increase on April 4th, when the blue-shifted area also reaches a maximum. The FIP bias then slowly decreases over the 
subsequent 5 days. The increase on the 4th is not dramatic, but it is about 50\% larger than measured on the 1st. This is potentially indicative of SEPs, or
is at least consistent with SEP activity, since they show a higher FIP bias in-situ than is measured in the slow solar wind \citep{Reames2018}. 
   
    To measure the FIP bias we used the Si {\sc x} 258.375\,\AA\ to S {\sc x} 264.223\,\AA\, line ratio, modified for temperature and density effects. We
measure the electron density using the Fe {\sc xiii} 202.044/203.826 diagnostic ratio. The density is then used to compute contribution functions for a collection of
spectral lines from low ($<$ 10\,eV) FIP elements observed by EIS. In this case we used lines of Fe covering a broad range of temperatures (0.52 -- 2.75\,MK). 
These are then used to compute the differential emission measure (DEM) distribution
from the observed intensities. With the DEM established, we can model the intensity of the high ($>$ 10\,eV) FIP S line. Since we used low FIP elements to derive the DEM, the calculated S {\sc x} 264.223\,\AA\, intensity will be too large if the FIP effect is operating. The ratio of computed to observed S {\sc x} 264.223\,\AA\, intensity is the FIP bias. This is a well established technique and the methodlogy we follow has been described in detail by \cite{Brooks2011} and \cite{Brooks2015}. We used the CHIANTI v.8 database \citep{Dere1997, DelZanna2015} to collect the atomic data for the contribution functions assuming the photospheric abundances
of \cite{Grevesse2007}. For the numerical computation of the DEM, we used the MCMC (Markov-Chain Monte Carlo) code in the PINTofALE software package \citep{Kashyap1998,Kashyap2000}. We also used the revised radiometric calibration of \cite{DelZanna2013} to take some account of the evolution of the EIS sensitivity.

We also show the average Doppler velocity and non-thermal velocity in the same box in this period. Both increase significantly in magnitude from the 1st to the 4th. There will be a line-of-sight component in the Doppler velocity but not in the non-thermal velocity which increases by more than a factor of 3.
    
%
    
   % The SEP events are frequent and dynamic during our time interval. We know that the active region is not flaring, so we looked for other dynamical behaviour. We looked first at the dynamics of the blue-shifted region.   We can do this as the EIS `rasters' are built up by moving the mirror which moves the position of the slit from right to left. Hence the `x' axes is a measure of time as well as spatial information.  Figure ~\ref{dynamics} shows an example of the line of Doppler shift extracted from the raster and plotted with time. Time is from right to left in the image. There are very frequent small variations across the line chosen with changes occurring on minutes timescales. The errors on the Doppler velocity are approximately 3.5 km/s \citep[see Appendix of ][]{Brooks2020}, so some of the rapid fluctuations are at the edge of the measurement capability of EIS. However the consistency of the measurement across all rasters and in different locations, supports the conclusion that the blue-shifted regions are dynamic. We performed the same analysis in the AR core and found that it shows fewer fluctuations. 
    
  %   \begin{figure*}
 %  \centering
  % \includegraphics[scale=0.6]{raster_dynamics1.pdf}
   %%%\includegraphics{empty.eps}
   %%%\includegraphics{empty.eps}
   %\caption{Top panel: EIS Doppler velocity map taken at 13:15UT on April 4. The black line marks the region of Doppler velocity we explore. Bottom panel: the time variation of the Doppler velocity.  }
    %          \label{dynamics}%
    % \end{figure*}
    
    In this section we have explored the active region, and its general characteristics, and we summarise them below.
    \begin{itemize}
\item  The AR 
%is the only active area on the disk during the period of the type III bursts and it
is a radio source and its dynamical evolution coincides with the evolution of the peak emission frequency of the dominant type III radio storm observed by PSP at this time.
\item  The active region has just emerged and as it evolves the magnetic field lines expand. In particular, at the edges of the active region.
\item  The edges of the active region both show increased Doppler velocities, increasing areas of upflows and increasing magnitude of upflows and non-thermal velocity between the 1st and 4th April. 
\item The active region does not flare or have jets. 

    \end{itemize}
    We conclude that the active region 
    %could be a source of the 
    is the most likely source of the energetic electron beams causing the type III radio storm, and more precisely, that the extended blue-shifted region could be a source. We also conclude that the changing nature of type III bursts (peak emission at a higher altitude or lower density region) must be related to the evolution and expansion of the active region. We investigate this further in the next section with high cadence observations. 
    
    \section{High cadence observations from Hinode EIS and comparison to PSP data}

    From the foregoing discussion, it is clear that the blue-shifted region is the most likely source of the SEPs within the studied active region. In this section, we analyse a high cadence EIS dataset to search for dynamical behaviour, similar to that which is seen in the PSP data, during a one hour period starting from 17:00UT on April 1. The EIS data consisted of a series of fast rasters over a small field-of-view, each of which took 41 seconds to complete. As already noted, the type III storm consists of continuous small and frequency localized bursts, and these are repeating on timescales of minutes.
    %bursts happen continuously in the PSP data \textbf{with a }so 
    This is another property that may help discriminate the source region.
    
\begin{figure*}
   \centering
   \includegraphics[scale=0.5]{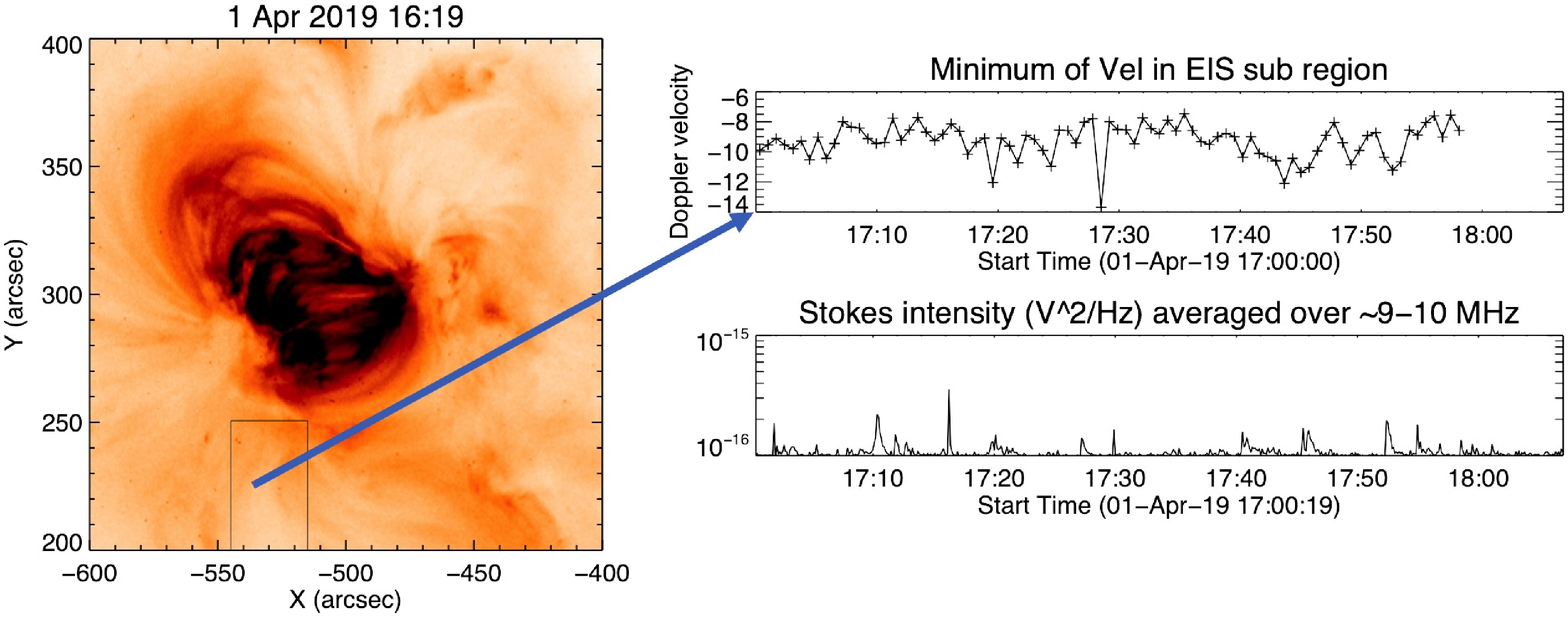}
   \caption{Left panel: AIA 193\,\AA\, image with the field-of-view of the EIS fast scan overlaid. Within the EIS raster we chose a small region of interest focused in the blue-shifted region. A blue arrow highlights this region of interest. Right panel, Top: variation of the Doppler velocity during the one hour raster. Bottom: variations in the PSP data and the frequency of the type III bursts for the same time interval. }
              \label{eis_psp}%
    \end{figure*}

    In the right panel of Figure~\ref{eis_psp}, the EIS Doppler velocity in the blue-shifted region is shown above, and the PSP data is shown below for the one hour time period. The Doppler velocities in this region are showing small but continuous variations, also on timescales of minutes. This is consistent with the nature of the type III bursts, and also with analysis carried out by \cite{inaki}. \cite{inaki} also found that the blue-shifted outflow regions showed transient blue wing enhancements within the 5 minute cadence of the their observations. The errors on the Hinode EIS Doppler velocity measurements are on the order of a few km/s, which means these fluctuations are on the edge of detectability for these measurements.

\begin{figure*}
   \centering
   \includegraphics[scale=0.6]{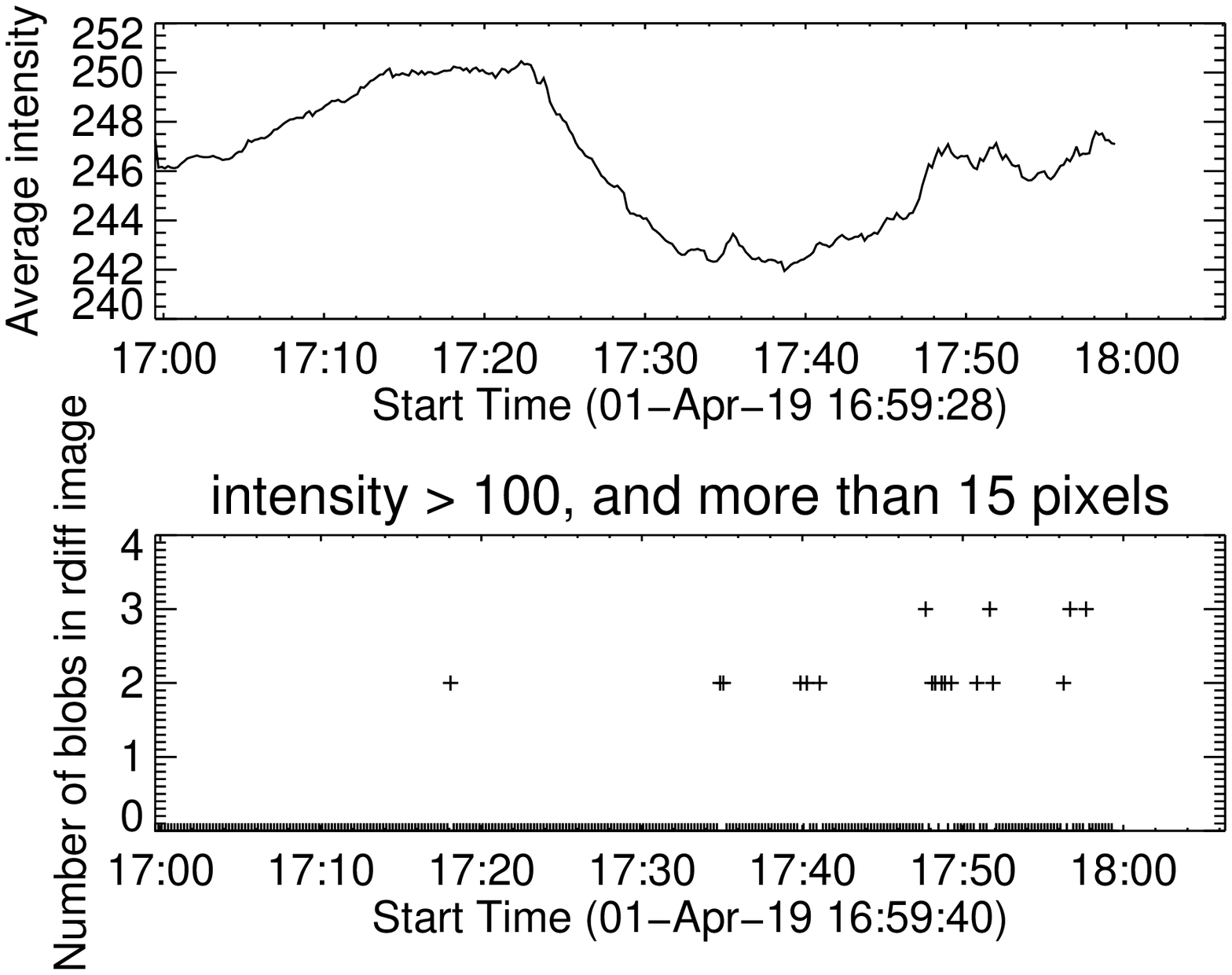}
   \caption{Top panel: lightcurve of the AIA intensity in the 193\,AA\, filter. Bottom panel: the number of brightenings in the AIA running difference movie that have a 'blob' size with an area more than 15 pixels and an intensity enhancement of $>$ 100\,DN. }
              \label{core_brightening}%
    \end{figure*}

     We investigated the properties of the AR core to see if it could potentially be the source through small-scale brightenings. We analysed the variability of the AR core in AIA data during the same one hour time period as the EIS high cadence scan. We already ruled out significant flaring during this time, but it is important to check for any small scale micro-flaring, so we produced a running difference movie to highlight any brightenings that occur in the core. We then extracted the number of brightenings which were defined as  having an area >  15 AIA pixels and have an intensity enhancement compared to the previous image of at least 100\,DN. Figure~\ref{core_brightening} shows the light curve of the whole active region. The longer timescale changes that are seen over $\approx$ 30 minutes are due to the intensity increase of `new' loops. The plot below shows when and how many brightenings occur in the core, and we find that they do not happen consistently during the whole hour period. We also searched in the XRT data for jets, and found none. Hence we infer that the fluctuations seen in the AR core are not viable as sources of the continuous type III bursts seen by PSP. 

\section{Conclusions}

We have analysed the behaviour of AR 12737 during the time period of 31 March - 6th April 
%--7 April 
2019 around perihelion (April 4) of the second encounter of PSP.
During this time, on a backdrop of larger, more impulsive type III bursts, PSP/FIELDS detected numerous small type III bursts constituting a radio noise storm. These type III bursts:
\begin{itemize}
    \item were rapid and persistent during the time interval,
    %\item increased in intensity by at least one order of magnitude from 1 to 4 April,  
    \item exhibited a decreasing peak frequency indicating a source altitude which is climbing in time, or a source region which is becoming more rarefied with time,
    \item both such possibilities are 
    %radio signatures 
    consistent with 
    % of 
    an expanding source region becoming more open to the solar wind. 
\end{itemize}

%AR 12737 is the only radio source on the Sun during this time period. 
AR 12737 is the most probable candidate source region for this type III noise storm. It is seen to emerge near the east limb at the same time as the radio noise storm develops. Between 1 and 4 April, as the noise storm evolved as described above, this active region also showed significant changes:
\begin{itemize}
    \item the area of the blue-shifted outflow region increased by an order of magnitude,
    \item the FIP bias increased in the blue-shifted region by a significant amount \citep[consistent with an increase in SEPs;][]{Reames2018},
    \item the whole active region expanded and, consequently, large-scale or expanding magnetic field lines anchored at the AR edge are more evident (including in the outflow region). 
    \item the magnitude of the Doppler velocity and the non-thermal velocity increases significantly as the active region expands in its first few days of formation.
\end{itemize}

The behaviour and changes in the AR during this time period are consistent with the source of the type III bursts being the blue-shifted outflow region. We also explored the dynamics of the outflow region which do show fluctuations on the time scale of the cadence of the observations. However these fluctuations are close to the measurement limit of the instrument, and provide a tantalising hint of dynamics. The high cadence aspect of these measurements is key to further understanding, and we encourage the future observing campaigns with PSP aim to have some fast cadence measurements.  

%We found that the Doppler velocities and FIP bias show small fluctuations on a time scale of minutes (the cadence limit of the EIS observations), and are varying continuously. These fluctuations are also close to the uncertainty limit of the EIS measurements, but provide additional evidence to support the suggestion that the blue-shifted outflow regions at the edges of AR 12737 are the source of the type III noise storm detected by PSP from 31 March - 4th April.  

In addition, the expansion of the blue-shifted area of the active region may offer an insight into the generation mechanism of type III storms. At least in this example, the decreasing peak frequency of the noise storm with time is suggestive of a scenario in which an expanding open field region allows the energetic electrons to more readily escape and thus produce their peak emission higher in the corona. Alternatively, the increasing open field region may be allowing more plasma to evacuate into interplanetary space, causing a rareification of the source region and hence a decreasing plasma frequency. 
The EIS composition measurements could be consistent with either scenario. Both an increase in escaping energetic particles and/or
source plasma would lead to a greater fraction of the total emission in the outflow areas we measured being contributed by higher FIP bias
plasma; therefore increasing the mean value. The decreasing trend in FIP bias after the peak on the 4th is consistent with a decrease in
the magnitude of the type III bursts.

\begin{acknowledgements}
      We acknowledge support from ISSI for the team 463, entitled 'Exploring The Solar Wind In Regions Closer Than Ever Observed Before'. 
      The work of DHB was performed under contract to the Naval Research Laboratory and was funded by the NASA Hinode program. LH is grateful for SNSF support.
      CHM acknowledges financial support from the Argentine grants PICT 2016-0221 (ANPCyT) and UBACyT 20020170100611BA (UBA). CHM is a member of the Carrera del Investigador Cient\'\i fico of the Consejo Nacional de Investigaciones Cient\'\i ficas y T\'ecnicas (CONICET).
      Hinode is a Japanese mission developed and launched by ISAS/ JAXA, with NAOJ as domestic partner and NASA and STFC (UK) as 
      international partners. It is operated by these agencies in cooperation with ESA and NSC (Norway). 
     This scientific work makes use of the Murchison Radio-astronomy Observatory (MRO), operated by the Commonwealth Scientific and Industrial Research Organisation (CSIRO). We acknowledge the Wajarri Yamatji people as the traditional owners of the Observatory site. Support for the operation of the MWA is provided by the Australian Government's National Collaborative Research Infrastructure Strategy (NCRIS), under a contract to Curtin University administered by Astronomy Australia Limited. We acknowledge the Pawsey Supercomputing Centre, which is supported by the Western Australian and Australian Governments.
      The SDO data are courtesy of NASA/SDO and 
      the AIA, EVE, and HMI science teams. CHIANTI is a collaborative project involving George Mason University, the University of Michigan (USA), 
      University of Cambridge (UK) and NASA Goddard Space Flight Center (USA).  The FIELDS instrument on
    the Parker Solar Probe spacecraft was designed and developed under NASA contract NNN06AA01C.
      Contributions from S.T.B. were supported by NASA Headquarters under the NASA Earth and Space Science Fellowship Program Grant 80NSSC18K1201. RS acknowledges support from the Swiss National Science foundation (grant no. 200021\_175832).
      Contribution from K.B. were supported by Swiss National Science Foundation.
\end{acknowledgements}

  \bibliographystyle{aa} % style aa.bst
   \bibliography{references.bib} % your references Yourfile.bib

\begin{thebibliography}{42}
\expandafter\ifx\csname natexlab\endcsname\relax\def\natexlab#1{#1}\fi

\bibitem[{{Badman} {et~al.}(2020){Badman}, {Bale}, {Mart{\'\i}nez Oliveros},
  {Panasenco}, {Velli}, {Stansby}, {Buitrago-Casas}, {R{\'e}ville}, {Bonnell},
  {Case}, {Dudok de Wit}, {Goetz}, {Harvey}, {Kasper}, {Korreck}, {Larson},
  {Livi}, {MacDowall}, {Malaspina}, {Pulupa}, {Stevens}, \&
  {Whittlesey}}]{Badman2020}
{Badman}, S.~T., {Bale}, S.~D., {Mart{\'\i}nez Oliveros}, J.~C., {et~al.} 2020,
  \apjs, 246, 23

\bibitem[{{Bale} {et~al.}(2019){Bale}, {Badman}, {Bonnell}, {Bowen}, {Burgess},
  {Case}, {Cattell}, {Chandran}, {Chaston}, {Chen}, {Drake}, {de Wit},
  {Eastwood}, {Ergun}, {Farrell}, {Fong}, {Goetz}, {Goldstein}, {Goodrich},
  {Harvey}, {Horbury}, {Howes}, {Kasper}, {Kellogg}, {Klimchuk}, {Korreck},
  {Krasnoselskikh}, {Krucker}, {Laker}, {Larson}, {MacDowall}, {Maksimovic},
  {Malaspina}, {Martinez-Oliveros}, {McComas}, {Meyer-Vernet}, {Moncuquet},
  {Mozer}, {Phan}, {Pulupa}, {Raouafi}, {Salem}, {Stansby}, {Stevens}, {Szabo},
  {Velli}, {Woolley}, \& {Wygant}}]{Bale2019}
{Bale}, S.~D., {Badman}, S.~T., {Bonnell}, J.~W., {et~al.} 2019, \nat, 576, 237

\bibitem[{{Bale} {et~al.}(2016){Bale}, {Goetz}, {Harvey}, {Turin}, {Bonnell},
  {Dudok de Wit}, {Ergun}, {MacDowall}, {Pulupa}, {Andre}, {Bolton},
  {Bougeret}, {Bowen}, {Burgess}, {Cattell}, {Chandran}, {Chaston}, {Chen},
  {Choi}, {Connerney}, {Cranmer}, {Diaz-Aguado}, {Donakowski}, {Drake},
  {Farrell}, {Fergeau}, {Fermin}, {Fischer}, {Fox}, {Glaser}, {Goldstein},
  {Gordon}, {Hanson}, {Harris}, {Hayes}, {Hinze}, {Hollweg}, {Horbury},
  {Howard}, {Hoxie}, {Jannet}, {Karlsson}, {Kasper}, {Kellogg}, {Kien},
  {Klimchuk}, {Krasnoselskikh}, {Krucker}, {Lynch}, {Maksimovic}, {Malaspina},
  {Marker}, {Martin}, {Martinez-Oliveros}, {McCauley}, {McComas}, {McDonald},
  {Meyer-Vernet}, {Moncuquet}, {Monson}, {Mozer}, {Murphy}, {Odom},
  {Oliverson}, {Olson}, {Parker}, {Pankow}, {Phan}, {Quataert}, {Quinn},
  {Ruplin}, {Salem}, {Seitz}, {Sheppard}, {Siy}, {Stevens}, {Summers}, {Szabo},
  {Timofeeva}, {Vaivads}, {Velli}, {Yehle}, {Werthimer}, \&
  {Wygant}}]{2016SSRv..204...49B}
{Bale}, S.~D., {Goetz}, K., {Harvey}, P.~R., {et~al.} 2016, \ssr, 204, 49

\bibitem[{{Brooks} {et~al.}(2015){Brooks}, {Ugarte-Urra}, \&
  {Warren}}]{Brooks2015}
{Brooks}, D.~H., {Ugarte-Urra}, I., \& {Warren}, H.~P. 2015, Nature
  Communications, 6, 5947

\bibitem[{{Brooks} \& {Warren}(2011)}]{Brooks2011}
{Brooks}, D.~H. \& {Warren}, H.~P. 2011, \apjl, 727, L13

\bibitem[{{Brooks} \& {Warren}(2016)}]{Brooks2016}
{Brooks}, D.~H. \& {Warren}, H.~P. 2016, \apj, 820, 63

\bibitem[{{Brooks} {et~al.}(2020){Brooks}, {Winebarger}, {Savage}, {Warren},
  {De Pontieu}, {Peter}, {Cirtain}, {Golub}, {Kobayashi}, {McIntosh},
  {McKenzie}, {Morton}, {Rachmeler}, {Testa}, {Tiwari}, \&
  {Walsh}}]{Brooks2020}
{Brooks}, D.~H., {Winebarger}, A.~R., {Savage}, S., {et~al.} 2020, \apj, 894,
  144

\bibitem[{{Culhane} {et~al.}(2007){Culhane}, {Harra}, {James}, {Al-Janabi},
  {Bradley}, {Chaudry}, {Rees}, {Tandy}, {Thomas}, {Whillock}, {Winter},
  {Doschek}, {Korendyke}, {Brown}, {Myers}, {Mariska}, {Seely}, {Lang}, {Kent},
  {Shaughnessy}, {Young}, {Simnett}, {Castelli}, {Mahmoud}, {Mapson-Menard},
  {Probyn}, {Thomas}, {Davila}, {Dere}, {Windt}, {Shea}, {Hagood}, {Moye},
  {Hara}, {Watanabe}, {Matsuzaki}, {Kosugi}, {Hansteen}, \& {Wikstol}}]{eis}
{Culhane}, J.~L., {Harra}, L.~K., {James}, A.~M., {et~al.} 2007, \solphys, 243,
  19

\bibitem[{{Del Zanna}(2013)}]{DelZanna2013}
{Del Zanna}, G. 2013, Astron. Astrophys., 555, A47

\bibitem[{{Del Zanna} {et~al.}(2011){Del Zanna}, {Aulanier}, {Klein}, \&
  {T{\"o}r{\"o}k}}]{DelZanna2011}
{Del Zanna}, G., {Aulanier}, G., {Klein}, K.~L., \& {T{\"o}r{\"o}k}, T. 2011,
  \aap, 526, A137

\bibitem[{{Del Zanna} {et~al.}(2015){Del Zanna}, {Dere}, {Young}, {Landi}, \&
  {Mason}}]{DelZanna2015}
{Del Zanna}, G., {Dere}, K.~P., {Young}, P.~R., {Landi}, E., \& {Mason}, H.~E.
  2015, Astron. Astrophys., 582, A56

\bibitem[{{D{\'e}moulin} {et~al.}(1997){D{\'e}moulin}, {Bagal{\' a}},
  {Mandrini}, {H{\'e}noux}, \& {Rovira}}]{Demoulin97}
{D{\'e}moulin}, P., {Bagal{\' a}}, L.~G., {Mandrini}, C.~H., {H{\'e}noux},
  J.~C., \& {Rovira}, M.~G. 1997, \aap, 325, 305

\bibitem[{{Dere} {et~al.}(1997){Dere}, {Landi}, {Mason}, {Monsignori Fossi}, \&
  {Young}}]{Dere1997}
{Dere}, K.~P., {Landi}, E., {Mason}, H.~E., {Monsignori Fossi}, B.~C., \&
  {Young}, P.~R. 1997, Astron. Astrophys. Supp. Ser., 125, 149

\bibitem[{{Fox} {et~al.}(2016){Fox}, {Velli}, {Bale}, {Decker}, {Driesman},
  {Howard}, {Kasper}, {Kinnison}, {Kusterer}, {Lario}, {Lockwood}, {McComas},
  {Raouafi}, \& {Szabo}}]{2016SSRv..204....7F}
{Fox}, N.~J., {Velli}, M.~C., {Bale}, S.~D., {et~al.} 2016, \ssr, 204, 7

\bibitem[{{Green} {et~al.}(2002){Green}, {L{\'o}pez fuentes}, {Mandrini},
  {D{\'e}moulin}, {Van Driel-Gesztelyi}, \& {Culhane}}]{Green02}
{Green}, L.~M., {L{\'o}pez fuentes}, M.~C., {Mandrini}, C.~H., {et~al.} 2002,
  \solphys, 208, 43

\bibitem[{{Grevesse} {et~al.}(2007){Grevesse}, {Asplund}, \&
  {Sauval}}]{Grevesse2007}
{Grevesse}, N., {Asplund}, M., \& {Sauval}, A.~J. 2007, Space Sci. Rev., 130,
  105

\bibitem[{{Harra} {et~al.}(2008){Harra}, {Sakao}, {Mandrini}, {Hara}, {Imada},
  {Young}, {van Driel-Gesztelyi}, \& {Baker}}]{flows}
{Harra}, L.~K., {Sakao}, T., {Mandrini}, C.~H., {et~al.} 2008, \apjl, 676, L147

\bibitem[{{Horbury} {et~al.}(2020){Horbury}, {Woolley}, {Laker}, {Matteini},
  {Eastwood}, {Bale}, {Velli}, {Chandran}, {Phan}, {Raouafi}, {Goetz},
  {Harvey}, {Pulupa}, {Klein}, {Dudok de Wit}, {Kasper}, {Korreck}, {Case},
  {Stevens}, {Whittlesey}, {Larson}, {MacDowall}, {Malaspina}, \&
  {Livi}}]{2020ApJS..246...45H}
{Horbury}, T.~S., {Woolley}, T., {Laker}, R., {et~al.} 2020, \apjs, 246, 45

\bibitem[{{Kamio} {et~al.}(2010){Kamio}, {Hara}, {Watanabe}, {Fredvik}, \&
  {Hansteen}}]{kamio2010}
{Kamio}, S., {Hara}, H., {Watanabe}, T., {Fredvik}, T., \& {Hansteen}, V.~H.
  2010, \solphys, 266, 209

\bibitem[{{Kashyap} \& {Drake}(1998)}]{Kashyap1998}
{Kashyap}, V. \& {Drake}, J.~J. 1998, Astrophys. J., 503, 450

\bibitem[{{Kashyap} \& {Drake}(2000)}]{Kashyap2000}
{Kashyap}, V. \& {Drake}, J.~J. 2000, Bulletin of the Astronomical Society of
  India, 28, 475

\bibitem[{{Kasper} {et~al.}(2019){Kasper}, {Bale}, {Belcher}, {Berthomier},
  {Case}, {Chandran}, {Curtis}, {Gallagher}, {Gary}, {Golub}, {Halekas}, {Ho},
  {Horbury}, {Hu}, {Huang}, {Klein}, {Korreck}, {Larson}, {Livi}, {Maruca},
  {Lavraud}, {Louarn}, {Maksimovic}, {Martinovic}, {McGinnis}, {Pogorelov},
  {Richardson}, {Skoug}, {Steinberg}, {Stevens}, {Szabo}, {Velli},
  {Whittlesey}, {Wright}, {Zank}, {MacDowall}, {McComas}, {McNutt}, {Pulupa},
  {Raouafi}, \& {Schwadron}}]{2019Natur.576..228K}
{Kasper}, J.~C., {Bale}, S.~D., {Belcher}, J.~W., {et~al.} 2019, \nat, 576, 228

\bibitem[{{Krucker} {et~al.}(2011){Krucker}, {Kontar}, {Christe}, {Glesener},
  \& {Lin}}]{krucker}
{Krucker}, S., {Kontar}, E.~P., {Christe}, S., {Glesener}, L., \& {Lin}, R.~P.
  2011, \apj, 742, 82

\bibitem[{{Krupar} {et~al.}(2020){Krupar}, {Szabo}, {Maksimovic}, {Kruparova},
  {Kontar}, {Balmaceda}, {Bonnin}, {Bale}, {Pulupa}, {Malaspina}, {Bonnell},
  {Harvey}, {Goetz}, {Dudok de Wit}, {MacDowall}, {Kasper}, {Case}, {Korreck},
  {Larson}, {Livi}, {Stevens}, {Whittlesey}, \& {Hegedus}}]{Krupar2020}
{Krupar}, V., {Szabo}, A., {Maksimovic}, M., {et~al.} 2020, \apjs, 246, 57

\bibitem[{{Leblanc} {et~al.}(1998){Leblanc}, {Dulk}, \&
  {Bougeret}}]{1998SoPh..183..165L}
{Leblanc}, Y., {Dulk}, G.~A., \& {Bougeret}, J.-L. 1998, \solphys, 183, 165

\bibitem[{{Lonsdale} {et~al.}(2009){Lonsdale}, {Cappallo}, {Morales}, {Briggs},
  {Benkevitch}, {Bowman}, {Bunton}, {Burns}, {Corey}, {Desouza}, {Doeleman},
  {Derome}, {Deshpande}, {Gopala}, {Greenhill}, {Herne}, {Hewitt}, {Kamini},
  {Kasper}, {Kincaid}, {Kocz}, {Kowald}, {Kratzenberg}, {Kumar}, {Lynch},
  {Madhavi}, {Matejek}, {Mitchell}, {Morgan}, {Oberoi}, {Ord},
  {Pathikulangara}, {Prabu}, {Rogers}, {Roshi}, {Salah}, {Sault}, {Shankar},
  {Srivani}, {Stevens}, {Tingay}, {Vaccarella}, {Waterson}, {Wayth}, {Webster},
  {Whitney}, {Williams}, \& {Williams}}]{Lonsdale2009}
{Lonsdale}, C.~J., {Cappallo}, R.~J., {Morales}, M.~F., {et~al.} 2009, IEEE
  Proceedings, 97, 1497

\bibitem[{{Mandrini} {et~al.}(2015){Mandrini}, {Baker}, {D{\'e}moulin},
  {Cristiani}, {van Driel-Gesztelyi}, {Vargas Dom{\'\i}nguez}, {Nuevo},
  {V{\'a}squez}, \& {Pick}}]{Mandrini15}
{Mandrini}, C.~H., {Baker}, D., {D{\'e}moulin}, P., {et~al.} 2015, \apj, 809,
  73

\bibitem[{{Mariska}(1994)}]{mariska}
{Mariska}, J.~T. 1994, \apj, 434, 756

\bibitem[{{Mohan} \& {Oberoi}(2017)}]{Mohan2017}
{Mohan}, A. \& {Oberoi}, D. 2017, \solphys, 292, 168

\bibitem[{{Morioka} {et~al.}(2015){Morioka}, {Miyoshi}, {Iwai}, {Kasaba},
  {Masuda}, {Misawa}, \& {Obara}}]{morioka2015}
{Morioka}, A., {Miyoshi}, Y., {Iwai}, K., {et~al.} 2015, \apj, 808, 191

\bibitem[{{Mulay} {et~al.}(2019){Mulay}, {Sharma}, {Valori}, {V{\'a}squez},
  {Del Zanna}, {Mason}, \& {Oberoi}}]{Mulay_Sharma_2019}
{Mulay}, S.~M., {Sharma}, R., {Valori}, G., {et~al.} 2019, \aap, 632, A108

\bibitem[{{Pesnell} {et~al.}(2012){Pesnell}, {Thompson}, \&
  {Chamberlin}}]{Pesnell2012}
{Pesnell}, W.~D., {Thompson}, B.~J., \& {Chamberlin}, P.~C. 2012, \solphys,
  275, 3

\bibitem[{{Pulupa} {et~al.}(2020){Pulupa}, {Bale}, {Badman}, {Bonnell}, {Case},
  {de Wit}, {Goetz}, {Harvey}, {Hegedus}, {Kasper}, {Korreck},
  {Krasnoselskikh}, {Larson}, {Lecacheux}, {Livi}, {MacDowall}, {Maksimovic},
  {Malaspina}, {Mart{\'\i}nez Oliveros}, {Meyer-Vernet}, {Moncuquet},
  {Stevens}, \& {Whittlesey}}]{pulupa}
{Pulupa}, M., {Bale}, S.~D., {Badman}, S.~T., {et~al.} 2020, \apjs, 246, 49

\bibitem[{{Pulupa} {et~al.}(2017){Pulupa}, {Bale}, {Bonnell}, {Bowen},
  {Carruth}, {Goetz}, {Gordon}, {Harvey}, {Maksimovic},
  {Mart{\'\i}nez-Oliveros}, {Moncuquet}, {Saint-Hilaire}, {Seitz}, \&
  {Sundkvist}}]{2017JGRA..122.2836P}
{Pulupa}, M., {Bale}, S.~D., {Bonnell}, J.~W., {et~al.} 2017, Journal of
  Geophysical Research (Space Physics), 122, 2836

\bibitem[{{Reames}(2017)}]{reames}
{Reames}, D.~V. 2017, {Solar Energetic Particles}, Vol. 932

\bibitem[{{Reames}(2018)}]{Reames2018}
{Reames}, D.~V. 2018, Space. Sci. Rev., 214, 61

\bibitem[{{Reid} \& {Ratcliffe}(2014)}]{hamish}
{Reid}, H. A.~S. \& {Ratcliffe}, H. 2014, Research in Astronomy and
  Astrophysics, 14, 773

\bibitem[{{Sterling} \& {Moore}(2020)}]{alphonse}
{Sterling}, A.~C. \& {Moore}, R.~L. 2020, \apjl, 896, L18

\bibitem[{{Tingay} {et~al.}(2013){Tingay}, {Goeke}, {Bowman}, {Emrich}, {Ord},
  {Mitchell}, {Morales}, {Booler}, {Crosse}, {Wayth}, {Lonsdale}, {Tremblay},
  {Pallot}, {Colegate}, {Wicenec}, {Kudryavtseva}, {Arcus}, {Barnes},
  {Bernardi}, {Briggs}, {Burns}, {Bunton}, {Cappallo}, {Corey}, {Deshpande},
  {Desouza}, {Gaensler}, {Greenhill}, {Hall}, {Hazelton}, {Herne}, {Hewitt},
  {Johnston-Hollitt}, {Kaplan}, {Kasper}, {Kincaid}, {Koenig}, {Kratzenberg},
  {Lynch}, {Mckinley}, {Mcwhirter}, {Morgan}, {Oberoi}, {Pathikulangara},
  {Prabu}, {Remillard}, {Rogers}, {Roshi}, {Salah}, {Sault}, {Udaya-Shankar},
  {Schlagenhaufer}, {Srivani}, {Stevens}, {Subrahmanyan}, {Waterson},
  {Webster}, {Whitney}, {Williams}, {Williams}, \& {Wyithe}}]{Tingay2013}
{Tingay}, S.~J., {Goeke}, R., {Bowman}, J.~D., {et~al.} 2013, \pasa, 30, e007

\bibitem[{{Ugarte-Urra} \& {Warren}(2011)}]{inaki}
{Ugarte-Urra}, I. \& {Warren}, H.~P. 2011, \apj, 730, 37

\bibitem[{{Wayth} {et~al.}(2018){Wayth}, {Tingay}, {Trott}, {Emrich},
  {Johnston-Hollitt}, {McKinley}, {Gaensler}, {Beardsley}, {Booler}, {Crosse},
  {Franzen}, {Horsley}, {Kaplan}, {Kenney}, {Morales}, {Pallot}, {Sleap},
  {Steele}, {Walker}, {Williams}, {Wu}, {Cairns}, {Filipovic}, {Johnston},
  {Murphy}, {Quinn}, {Staveley-Smith}, {Webster}, \& {Wyithe}}]{MWA_Phase2}
{Wayth}, R.~B., {Tingay}, S.~J., {Trott}, C.~M., {et~al.} 2018, \pasa, 35, 33

\bibitem[{{Zlotnik}(1981)}]{1981A&A...101..250Z}
{Zlotnik}, E.~I. 1981, \aap, 101, 250

\end{thebibliography}

%\begin{thebibliography}{}
%\bibitem[Reames, 2017]{reames} Reames, D. 2017,
  %   Lecture Notes in Physics, Vol. 932 

%\end{thebibliography}

   \end{document}